# Scalable hierarchical parallel algorithm for the solution of super large-scale sparse linear equations

R. Xu [a], B. Liu [a,*], Y. Dong [b,*]

[a] AML, CNMM, Department of Engineering Mechanics, Tsinghua University, Beijing 100084, China
[b] Department of Computer Science and Technology,, Tsinghua University, Beijing 100084, China

*Corresponding author.
*Tel*: 86-10-6278-6194; *Fax*: 86-10-6278-1824; *E-mail address*:
liubin@tsinghua.edu.cn (B. Liu), dongyuan@tsinghua.edu.cn (Y. Dong).

**ABASTRACT**

The parallel linear equations solver capable of effectively using 1000+ processors becomes the bottleneck of large-scale implicit engineering simulations. In this paper, we present a new hierarchical parallel master-slave-structural iterative algorithm for the solution of super large-scale sparse linear equations in distributed memory computer cluster. Through alternatively performing global equilibrium computation and local relaxation, our proposed algorithm will reach the specific accuracy requirement in a few of iterative steps. Moreover, each set/slave-processor majorly communicate with its nearest neighbors, and the transferring data between sets/slave-processors and master is always far below the set-neighbor communication. The corresponding algorithm for implicit finite element analysis has been implemented based on MPI library, and a super large 2-dimension square system of triangle-lattice truss structure under random static loads is simulated with over one billion degrees of freedom and up to 2001 processors on "Exploration 100" cluster in Tsinghua University. The numerical experiments demonstrate that this algorithm has excellent parallel efficiency and high scalability, and it may have broad application in other implicit simulations.

**KEY WORDS**

Super large linear equations; Parallel computing; Finite element method; Iterative algorithm; Hierarchical

## 1. Introduction

Many numerical simulations in engineering and scientific fields involve solving large scale sparse linear equations, such as implicit finite element method (FEM) for structure analysis. Thanks to the advancement of computer technology, parallel computing with more than one thousand processor makes it possible to model super large-scale problems from hardware aspect. However, as far as we know, the efficient and scalable (up to 1000+ processors) parallel algorithm for solving sparse linear equations still lacks, which seriously restricts the relevant computational capability.

How to increase the efficiency and scalability of parallel algorithm on sparse linear equations has attracted research attentions for many years, and various schemes have been proposed [1-16]. For example, Saad et al. [7] investigated the influence the partition of parallel computing on its efficiency, and graph partitioning concept was adopted. Censor et al. [8] proposed a so-called component averaging approach, which simultaneously projects the current iteration onto all the system's hyperplanes, and is thus inherently parallel. Its convergence was demonstrated on image reconstruction problem. Amestoy et al. [12] designed a dynamic scheduling strategy to balance the workload of factorizing linear equations, and the numerical example on 64 and 128 processor demonstrated that the memory overhead and the factorization time were improved. Manguolu et al. [16] developed the PSPIKE hybrid sparse linear system solver, which has a better speed improvement compared with PARDISO and MUMPS on a distributed memory platform with 256 processors. Agullo et al. [13] proposed a hierarchical algorithmic approach, which provides coarse grain parallelism between nodes and fine grain parallelism within each node. On these basics, a lot of paralleling linear system solver is developed, such as SuperLU [17] MUMPS [18] PETSC [19] Hypre [20] and PSPIKE [16]. However, although these existing algorithms improve the parallel efficiency from different aspects, as far as we know, there is no one possessing the linear scalability when more than 512 processors are used.

In the area of finite element method, many scholars have also been working on improving the parallel efficiency of solution of linear equations by incorporating the features of realistic problems [21-31]. Among them, Law [21] proposed a parallel finite element method, which does not require the formation of global system equations and computes local quantities with each processor. Farhat and his co-workers [22, 23] proposed to partition the spatial domain into a set of totally disconnected sub-domains, and Lagrange multipliers are then introduced to enforce compatibility at the interface nodes. Tezduyar et al. [25] investigated a special preconditioning techniques to enhance the computational schemes. Parallel multi-grid algorithm [29-31] has been employed to solve mechanics problems with unstructured or structured finite element meshes by many researchers, and Adams et al. [32] studied a human vertebral body problem with half of a billion degrees of freedom (DOFs). Still, the parallel efficiency of all these parallel finite element methods can not be maintained for more than 512 processors. By the example of multi-grid method, it is reported that the communication accounts for up to 90% of the execution time for more than 32 processors [33].

Based on the previous studies, we can summarize various strategies to improve the parallel efficiency in solving sparse linear equations as follows,
- i) preconditioning;
- ii) balancing the computational load in each processor;
- iii) optimal partitioning;
- iv) reducing the inter-processor communication;
- v) reducing the number of required iterations for domain decomposition or hierarchical schemes.

Obviously, the last two aspects are more crucial in keeping the linear parallel scalability. If the number of required iterations, or the inter-processor communication for a processor, increases significantly with the number of processors (i.e., the problem scale), the corresponding parallel algorithm would not have good performance for super large-scale problem with more than one thousand processors.

Our approach is a hierarchical parallel master-slave-structural iterative method. The main idea is alternatively performing global equilibrium computation and local relaxation, and each computation corresponds to solving small-scale linear equations. The paper is organized as followed. Section 2 presents the derivation of the algorithm, including how to construct the upper-level equations and chosen of incremental mode for unknowns on the bottom-level. In section 3, various parallel characteristics of our algorithm, such as the convergence, scalability and parallel performance, are tested with several implicit finite element problems. The conclusions and discussions about superiority and potential extension of our algorithm are given in Section 4.

## 2. The hierarchical parallel algorithm of sparse linear equations solver

### 2.1 The derivation of the algorithm

The linear equations to be solved is

$$\sum_{j=1}^{N} \mathbf{k}_{ij} \cdot \mathbf{u}_j = \mathbf{p}_i, \; i = 1, 2, \cdots N, \qquad (1)$$

In this paper, we will borrow the language, even some ideas, from the FEM for solid mechanics. But it should be pointed out that the proposed algorithm can be directly used without any revision to solve other sparse linear equations via parallel computing. Assume that a FEM system has $N$ nodes, and the number of degrees of freedom for each node is $s$, Therefore, $[\mathbf{k}_{ij}]$ in Eq.(1) is the global stiffness matrix with dimension $sN \times sN$, $[\mathbf{p}_i]$ and $[\mathbf{u}_i]$ are force matrix and unknown displacement matrix with dimension $sN \times 1$, respectively.

The global stiffness $[\mathbf{k}_{ij}]$ is assumed to be symmetrical and positive definite, otherwise a preconditioning can be implemented through multiplying both sides of Eq.(1) by its transpose, which yields a new set of linear equations with symmetrical and positive definite stiffness matrix.

Determining the solution of the linear equations (1) is equivalent to seeking the minimum of the following total energy with respect to $\mathbf{u}_i$,

$$\Pi = \frac{1}{2} \sum_{i,j=1}^{N} \mathbf{u}_i \cdot \mathbf{k}_{ij} \cdot \mathbf{u}_j - \sum_{i=1}^{N} \mathbf{u}_i \cdot \mathbf{p}_i . \qquad (2)$$

Hereafter in this paper, we will design an iterative scheme to ensure the total energy $\Pi$ to decrease in every iteration and finally approach a convergent value.

For parallel computation, all nodes are decomposed into $M$ sets (or sub-domains in previous literatures) according to the principle that the nodes in each set should have least connections with the nodes outside this set, as shown in Fig. 1. For example, a meshed solid structure is divided into sets/sub-domains corresponding to the location. In our algorithm, the boundary line cuts through elements, and no more additional restrictions are introduced to enforce the continuity at the set/sub-domain interfaces. Then these $M$ sets are assigned to $M$ processors.

**Figure 1**

In each set, a small number of incremental displacement modes are introduced. For example, if node $i$ belongs to $set_I$, its displacement is

$$\mathbf{u}_i = \mathbf{u}_i^{old} + \sum_{k=1}^{q} a_k^{(I)} \hat{\mathbf{u}}_{ki}^{(I)}, \qquad i \in set_I, \tag{3}$$

where $\mathbf{u}_i^{old}$ is the displacement after last iteration, $q$ is the number of incremental displacement modes, $\hat{\mathbf{u}}_{ki}^{(I)}$ is the $k^{th}$ incremental displacement mode for $set_I$ and $a_k^{(I)}$ is the corresponding coefficient to be determined. How to choose $\hat{\mathbf{u}}_{ki}^{(I)}$ will be discussed in Section 2.2. Through Eq. (3), a large number of DOFs ($sN/M$) for each set at the bottom fine level have been condensed into $q$ coefficient, or $q$ DOFs for the upper second level, which can be determined by solving smaller-scale linear equations at the upper level.

Substituting Eq. (3) into Eq. (2), the total energy can be expressed in terms of $a_k^{(I)}$ as,

$$\Pi = \frac{1}{2} \sum_{I,J=1}^{M} \sum_{\substack{i \in \text{Set}_I \\ j \in \text{Set}_J}} \left( \mathbf{u}_i^{old} + \sum_{k=1}^{q} a_k^{(I)} \hat{\mathbf{u}}_{ki}^{(I)} \right) \cdot \mathbf{k}_{ij} \cdot \left( \mathbf{u}_j^{old} + \sum_{l=1}^{q} a_l^{(J)} \hat{\mathbf{u}}_{lj}^{(J)} \right)$$
$$- \sum_{I=1}^{M} \sum_{i \in \text{Set}_I} \mathbf{p}_i \cdot \left( \mathbf{u}_i^{old} + \sum_{k=1}^{q} a_k^{(I)} \hat{\mathbf{u}}_{ki}^{(I)} \right) \tag{4}$$

The energy minimum requires

$$\frac{\partial \Pi}{\partial a_m^{(I)}} = \sum_{J=1}^{M} \sum_{\substack{i \in \text{Set}_I \\ j \in \text{Set}_J}} \hat{\mathbf{u}}_{mi}^{(I)} \cdot \mathbf{k}_{ij} \cdot \left( \mathbf{u}_j^{old} + \sum_{l=1}^{q} a_l^{(J)} \hat{\mathbf{u}}_{lj}^{(J)} \right) - \sum_{i \in \text{Set}_I} \mathbf{p}_i \cdot \hat{\mathbf{u}}_{mi}^{(I)}$$
$$= \sum_{J=1}^{M} \sum_{l=1}^{q} K_{a_m^{(I)} a_l^{(J)}} a_l^{(J)} - P_{a_m^{(I)}} \tag{5}$$
$$= 0$$

where

$$K_{a_m^{(I)} a_l^{(J)}} = \sum_{\substack{i \in \text{Set}_I \\ j \in \text{Set}_J}} \hat{\mathbf{u}}_{mi}^{(I)} \cdot \mathbf{k}_{ij} \cdot \hat{\mathbf{u}}_{lj}^{(J)}, \tag{6}$$

$$P_{a_m^{(I)}} = \sum_{i \in \text{Set}_I} \hat{\mathbf{u}}_{mi}^{(I)} \cdot \left( \mathbf{p}_i - \sum_{J=1}^{M} \sum_{j \in \text{Set}_J} \mathbf{k}_{ij} \cdot \mathbf{u}_j^{old} \right) \tag{7}$$

are the components of the stiffness matrix and the force matrix at the upper level, respectively. From Eq. (5), the corresponding linear equations at this level can be obtained as

$$\sum_{J=1}^{M}\sum_{l=1}^{q} K_{a_m^{(I)} a_l^{(J)}} a_l^{(J)} = P_{a_m^{(I)}}, \quad m=1,2,\cdots,q, \quad I=1,2,\cdots,M. \tag{8}$$

The scale of Eq. (8) is much smaller than that of Eq. (1), and can be solved within an additional processor, which is called the *second-level processor* hereafter.

Once $a_l^{(J)}$ are determined, $\mathbf{u}_i$ can be updated according to Eq. (3). It should be pointed out that comparing with $\mathbf{u}_i^{old}$, $\mathbf{u}_i$ corresponds to lower energy $\prod$, therefore it is a better approximate solution.

Since the incremental displacement for each set is represented with only $q$ DOFs, one time solution must be very rigid, and iterations with adjusting incremental displacement modes $\hat{\mathbf{u}}_{ki}^{(I)}$ are needed.

## 2.2 Selection of incremental displacement modes for each set

Two groups of incremental displacement modes are introduced: the first one should be capable of characterizing the deformation to make the global equilibrium satisfied at the coarse and upper level quickly; the second one is focused on the local equilibrium of each set and its neighboring sets.

For example, if the unknown variables are two dimensional displacements of nodes, the first group of incremental displacement modes should include *translational motion modes*

$$\hat{\mathbf{u}}_{1i}^{(I)} = \begin{bmatrix} 1 \\ 0 \end{bmatrix}, \quad \hat{\mathbf{u}}_{2i}^{(I)} = \begin{bmatrix} 0 \\ 1 \end{bmatrix} \tag{9}$$

and *constant displacement gradient modes*

$$\hat{\mathbf{u}}_{3i}^{(I)} = \begin{bmatrix} x_i - x^{(I)} \\ 0 \end{bmatrix}, \quad \hat{\mathbf{u}}_{4i}^{(I)} = \begin{bmatrix} y_i - y^{(I)} \\ 0 \end{bmatrix}$$
$$\hat{\mathbf{u}}_{5i}^{(I)} = \begin{bmatrix} 0 \\ x_i - x^{(I)} \end{bmatrix}, \quad \hat{\mathbf{u}}_{6i}^{(I)} = \begin{bmatrix} 0 \\ y_i - y^{(I)} \end{bmatrix} \tag{10}$$

where $x_i, y_i$ are the coordinates of node $i$, and $x^{(I)}, y^{(I)}$ are the coordinates of representing node of set $I$.

Since the above incremental displacement modes of the first group are fixed and limited, the second group of incremental displacement modes are needed, which are adaptively adjustable based on local equilibrium. Two types of local relaxation are adopted in this paper as shown in Fig. 2: one is fixing the displacement of the outer region during the relaxation (see Fig. 2(b)); the other is keeping the forces from the outer region constant during the relaxation (see Fig. 2(c)). The specific formulae are given as follows.

*Local relaxation with fixed-displacement boundary*

Assume there are $n$ inner nodes belonging to $\text{Set}_I$, and $m$ outer nodes outside $\text{Set}_I$ with direct relation with inner nodes, as shown in Fig. 2(a). For $\text{Set}_I$, the local equilibrium equations are

$$\begin{bmatrix} \mathbf{k}^{\text{Inner}} & \mathbf{k}^{\text{Outer}} \end{bmatrix} \begin{bmatrix} \mathbf{u}^{\text{Inner}} \\ \mathbf{u}^{\text{Outer}} \end{bmatrix} = \mathbf{p}^{\text{Inner}}. \tag{11}$$

where $\mathbf{k}^{\text{Inner}}$ and $\mathbf{k}^{\text{Outer}}$ are the stiffness matrixes for inner nodes and outer nodes, respectively, $\mathbf{p}^{\text{Inner}}$ is force matrix of $\text{Set}_I$, and the components of these matrixes are the same as those in the global matrixes in Eq.(1). $\mathbf{u}^{\text{Inner}}$ and $\mathbf{u}^{\text{Outer}}$ are the displacement vectors of inner nodes and outer nodes. According to Eq. (11), local relaxation with fixed-displacement boundary can be implemented by fixing the last approximate displacement $\mathbf{u}^{\text{Outer-old}}$ of outer nodes, and determining the relaxation displacement of inner nodes $\mathbf{u}^{\text{Inner-RD}}$ as

$$\mathbf{k}^{\text{Inner}} \cdot \mathbf{u}^{\text{Inner-RD}} = \mathbf{p}^{\text{Inner}} - \mathbf{k}^{\text{Outer}} \cdot \mathbf{u}^{\text{Outer-old}}, \tag{12}$$

which is essentially an equilibrium equation. The incremental displacement mode for displacement-controlled relaxation is then obtained as

$$\hat{\mathbf{u}}_{7i}^{(I)} = \mathbf{u}_i^{\text{Inner-RD}} - \mathbf{u}_i^{\text{Inner-old}}. \tag{13}$$

The following displacement-controlled relaxation incremental modes with spatial gradient are also introduced to accelerate convergence,

$$\begin{aligned} \hat{\mathbf{u}}_{8i}^{(I)} &= (x_i - x^{(I)})\left(\mathbf{u}_i^{\text{Inner-RD}} - \mathbf{u}_i^{\text{Inner-old}}\right) \\ \hat{\mathbf{u}}_{9i}^{(I)} &= (y_i - y^{(I)})\left(\mathbf{u}_i^{\text{Inner-RD}} - \mathbf{u}_i^{\text{Inner-old}}\right) \end{aligned}. \tag{14}$$

*Local relaxation with fixed-force boundary*

Figure 2c schematically shows the isolated $\text{Set}_I$ subjected to fixed boundary force $\bar{\mathbf{p}}$. Obviously, the local stiffness and force matrix of $\text{Set}_I$ are different from their counterparts in the displacement-controlled case. The difference of the stiffness is denoted as $\Delta \mathbf{k}$, and according to traditional FEM,

$$\Delta \mathbf{k}_{ij} = \begin{cases} \sum_{l \notin \text{Set}_I} \mathbf{k}_{il}^{\text{Outer}} & i = j \\ 0 & i \neq j \end{cases}, \quad i \in \text{Set}_I, \; j \in \text{Set}_I \tag{15}$$

Equilibrium Equation (11) for determining the relaxation displacement of inner nodes $\mathbf{u}^{\text{Inner-RF}}$ becomes

$$\left(\mathbf{k}^{\text{Inner}} + \Delta \mathbf{k}\right) \cdot \mathbf{u}^{\text{Inner-RF}} = \mathbf{p}^{\text{Inner}} - \mathbf{k}^{\text{Outer}} \cdot \mathbf{u}^{\text{Outer-old}} + \Delta \mathbf{k} \cdot \mathbf{u}^{\text{Inner-old}} = \bar{\mathbf{p}}, \tag{16}$$

Similarly, the incremental displacement mode for force-controlled relaxation is

$$\hat{\mathbf{u}}_{10i}^{(I)} = \mathbf{u}_i^{\text{Inner-RF}} - \mathbf{u}_i^{\text{Inner-old}}. \tag{17}$$

and the corresponding relaxation incremental modes with spatial gradient are

$$\hat{\mathbf{u}}_{11i}^{(I)} = \left(x_i - x^{(I)}\right)\left(\mathbf{u}_i^{\text{Inner-RF}} - \mathbf{u}_i^{\text{Inner-old}}\right)$$
$$\hat{\mathbf{u}}_{12i}^{(I)} = \left(y_i - y^{(I)}\right)\left(\mathbf{u}_i^{\text{Inner-RF}} - \mathbf{u}_i^{\text{Inner-old}}\right). \quad (18)$$

To achieve the better relaxation of $\text{Set}_I$, the local relaxation can be implemented with a larger extended region of $\text{Set}_I$, as the region surrounded by dot dash line in Fig. 1. The larger relaxation region leads to faster convergence or less iteration, but more computation effort in the iterations.

**Figure 2**

Moreover, the fact should be noted that the stiffness matrixes for local relaxation in (12) and (16) remain unchanged during iterations, which can be used to reduce the computation effort if the direct serial linear equations solver is adopted, such as PARDISO package [11, 34] or direct sparse solver (DSS). In specific, we perform the factorization in the first iteration and keep the factorization result in internal memory. In the following iterations, we only use the stored factorization with different force vectors $\mathbf{p}$ to obtain the solution.

## 3. Numerical experiments

A procedure of the hierarchical sparse linear equations solver has been developed in Fortran 90 programming language and MPI 1.0, which is an open-source library specification for message-passing in distributed-memory computer cluster[35, 36].

Fig. 3 shows the flowchart of the algorithm. At the beginning, master process reads the entire linear equations or discrete system information, and sends to slave ones. After that, both master and slave processes start a loop until the stopping criterion is satisfied, and the stopping criterion here is chosen as the following 1-norm relative residual,

$$R = \frac{\left\|\sum_{j=1}^{N}\mathbf{k}_{ij}\cdot\mathbf{u}_j - \mathbf{p}_i\right\|_1}{\left\|\mathbf{p}_i\right\|_1} < \varepsilon = 5\times 10^{-6}. \quad (19)$$

In each loop, slave processes start with choosing incremental modes, and then call a single-CPU linear equations solver for local relaxation. The PARDISO package [11, 34] in Intel Math Kernel Library is used to solve problem (12) and (16). The upper-level equation (5) is constructed in the slave processes, and is solved in the master process also through a single-CPU linear equations solver. After receiving the coefficients of increment modes from master, slaves update the unknowns with formula (3) and calculate immediate relative residual $R$ of whole system. If the stopping criterion is satisfied, the loop will be broken. The source part running on the master and slave processes communicate with each other through MPI_SEND MPI_RECV MPI_GATHER and MPI_REDUCE functions.

**Figure 3**

## 3.1 Test problems and experimental environment

As shown in Fig. 4, we choose a series of 2-dimension square system with triangle-lattice truss structures as test problems. Each node in the test truss is connected with neighbor nodes by elastic rods. Different test problems have different scales or DOFs. To test the scalability and validity of the algorithm, the loading condition should be consistent and as complex as possible, therefore each node is assumed to subject to a force which is random both in magnitude and direction. The displacements of nodes can be determined by solving corresponding linear equations via standard FEM routines.

**Figure 4**

The parallel machine used to run our algorithm is "Exploration 100" cluster in Tsinghua University. The cluster has 740 computing nodes each with two six-core CPUs, for a total of 8800 processors/cores, interconnected with 40 Gigabit InfiniBand QDR Ethernet. Each computing node is a blade server with two Xeon X5670 CPUs at 2.93GHz and 32GB RAM. The storage system of "Exploration 100" cluster is composed of 22 data nodes for a total of 160TB, which are charged by LUSTRE parallel file system with 4 GB/s write speed. More specific details on the cluster are given in Table 1.

**Table 1**

## 3.2 Tests on the convergence and scalability

*Test problem 1: 1 billion global DOFs, 2000 sets, 500,000 DOFs/set*

In order to directly reveal the capacity of the proposed algorithm, the first test problem is a super large truss with one billion DOFs. The truss is evenly divided into 2000 sets for load balance, and each set has half of million DOFs. According to the algorithm, 2001 processors in "Exploration 100" cluster is used, one of which is used to run master process. Fig. 5 shows the variation of the relative residual $R$ as a function of the iterations number. Apparently, $R$ decreases steadily in a quasi-exponential way with the increasing number of iterations. If we choose (19) as the stopping criterion, the execution time is only 945.53 seconds for this super large problem with one billion DOFs, and the utilization rate of all processors exceeds 94% during the whole computation.

**Figure 5**

The number of iteration required (NIR) to meet specific accuracy requirement is a key in iterative algorithms. In the following, the dependence of NIR on the problem size and the number of sets in parallel computing is investigated.

*Test problem 2: Fixing the number of sets while increasing the DOFs in each set*

The 2-dimension square system is averagely divided into 64 sets, and the number of DOFs in each set ranges from 800 to 980,000. Four different random loading conditions are tested, and the average NIR versus the number of DOFs in each set is shown in Fig. 6. At very beginning the average NIR increases with the number of DOFs in each set, and then quickly converge to a fixed value, which implies the NIR is roughly independent of the number of DOFs in each set for large problems.

**Figure 6**

*Test problem 3: Fixing the DOFs in each set while increasing the number of sets*

The 2-dimension square system is averagely divided into 2 ~ 2000 sets, and the number of DOFs in each set is 500,000. Four different random loading conditions are tested, and the average NIR versus the number of sets is shown in Fig. 7(a). It is easy to find that NIR shows a rapid convergence as the number of sets increases. The corresponding elapsed time per iteration versus number of sets is also shown in Fig. 7(b). It can be observed that the elapsed time per iteration has a very slow growth rate with increasing set number, which demonstrates a good scalability of our algorithm.

**Figure 7**

### 3.3 Parallel performance evaluation

*Test problem 4: Fixing the number of global DOFs while increasing the number of sets*

We apply the solver scheme on two truss examples. The smaller one has eight million DOFs, while the larger one has 32 million DOFs. The parallel improvement of the algorithms is evaluated by analyzing the speed-up $S_n$ defined by the following formula

$$S_n = \frac{T_1}{T_n}, \tag{20}$$

which reflects how much the parallel algorithm is faster than the corresponding sequential algorithm. In formula (20), $n$ is the number of sets, and $T_n$ is the execution time of the parallel algorithm with $n$ sets. Linear speed-up $S_n = n$ is considered as the ideal state and has a very good scalability. Generally $T_1$ in formula (20) is the execution time running with a single set. However, to better demonstrate the performance of our iterative parallel solver, the benchmark test is alternatively chosen as the case with 4 sets with assigned speed-up value of 4.

The resulting curves of speed-up are presented in Fig. 8. It can be observed that our algorithm has an apparent advantage, and the speed up curves keep super linear improvement till 625 sets for the case with 8 million DOFs. One possible reason for the super linear speed-up is that the PARDISO package is not a strictly order-N linear equations solver, and the smaller scale problem needs much less time. We may imagine that if an order-N linear equations solver is used in our code, the parallel speed-up curve will approach the ideal line. It is also noted in Fig. 8 that as increasing the number of sets or decreasing DOFs in each set, the speed-up shows a rise first followed by a decline. This phenomenon can be explained by the structure of our algorithm. Overall this algorithm is implemented in a serial way: one step is solving local small-scale equations in slave processors, and the next step is solving the upper-level equations in the master processor. If the major time consumption occurs in the slave processors, increasing the number of sets/processors can reduce the scale of local equations and increase the efficiency, just like the increasing part of the speed-up curve. However, if the major time consumption occurs in the master processor, increasing the number of sets/processors will lead to larger scale equations in the master processor and lower efficiency. Hence, there theoretically exists the peak on the speed-up curve. Fortunately, this peak occurs later for the larger-scale problem, such as the case of 32 million DOFs in Fig. 8 without peak below 1024 sets.

**Figure 8**

## 4. Conclusions and Discussions

Through iteratively performing global equilibrium computation and local relaxation, a new hierarchical parallel algorithm for sparse linear equations is proposed in this paper, and it is demonstrated to be scalable and computationally effective for super-large implicit finite element analysis with up to one billion degrees of freedom and 2001 processors. Moreover, the following features are discussed or highlighted to extend its applications.

**(i) A hierarchical algorithm for super large-scale linear equations:** In this paper, the large-scale global linear equations are iteratively solved by solving small-scale equations on the upper and bottom levels, and the corresponding computation capacity has reached 1 billion DOFs. This strategy can be easily extended to the multiple-level algorithm, and the computation capacity can be further increased without obvious limit.

**(ii) Incremental modes for general sparse linear equations**: Although we used solid structures to demonstrate our algorithm, actually general symmetric linear equations are equivalent to a one-dimensional spring system. For example, the following linear equations with three unknowns

$$\begin{pmatrix} k_{11} & k_{12} & k_{13} \\ k_{21} & k_{22} & k_{23} \\ k_{31} & k_{32} & k_{33} \end{pmatrix} \begin{pmatrix} u_1 \\ u_2 \\ u_3 \end{pmatrix} = \begin{pmatrix} p_1 \\ p_2 \\ p_3 \end{pmatrix}, \quad (k_{ij} = k_{ji}) \qquad (21)$$

represent the equilibrium problem of a one dimensional spring system shown in Fig. 9. Once this equivalence is established, all treatments used on solid elements can be applied to general linear equations.

**(iii) Non-symmetric coefficient matrix**: As mentioned above, we can transfer the non-symmetric matrix into a symmetric one by pre-multiplication of the transpose of the matrix to Eq. (1),

$$\sum_{j=1}^{N} \sum_{m=1}^{N} \mathbf{k}_{mi} \cdot \mathbf{k}_{mj} \cdot \mathbf{u}_j = \sum_{m=1}^{N} \mathbf{k}_{mi} \cdot \mathbf{p}_m . \qquad (22)$$

where $\sum_{m=1}^{N} \mathbf{k}_{mi} \cdot \mathbf{k}_{mj} = \sum_{m=1}^{N} \mathbf{k}_{mj} \cdot \mathbf{k}_{mi}$. Then, (22) can be solved by our algorithm. The cell-sparse storage scheme [37] may be a good method to improve the efficiency of parallel pre-multiplication.

**(iv) Preprocessing and communication:** Different from the other iterative method, our algorithm does not need to call incomplete LU precondition at the beginning. According to our algorithm, each set/slave-processor majorly communicate with its nearest neighbors, and the transferring data between sets/slave-processors and master is in proportion to the number of chosen incremental modes, which is always far below the set-neighbor communication. Therefore, the intensive all-to-all communication is avoided. For message passing platforms, we can call the non-blocking communication routines to overlap the computation and communication, and the utilization rate of processors can be further improved.

**Figure 9**

## Acknowledgements

The authors are grateful for the support from National Natural Science Foundation of China (Grant Nos. 10732050, 90816006, 11090334, and 10820101048), and National Basic Research Program of China (973 Program) Grant No. 2007CB936803 and 2010CB832701. The computations on the "Exploration 100" cluster were supported by Tsinghua National Laboratory for Information Science and Technology. We would like to thank Dr. Jiao Lin for help with advices about parallel performance, and Prof. Wei Xue for help in the beginning of procedure development.

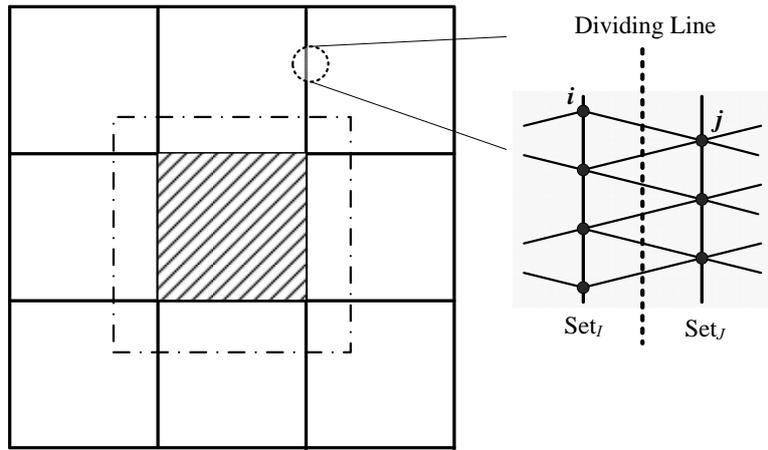

Fig. 1 A meshed 2-dimension solid structure is divided into sets/sub-domains assigned to slave processors, and the dividing line cuts through the elements.

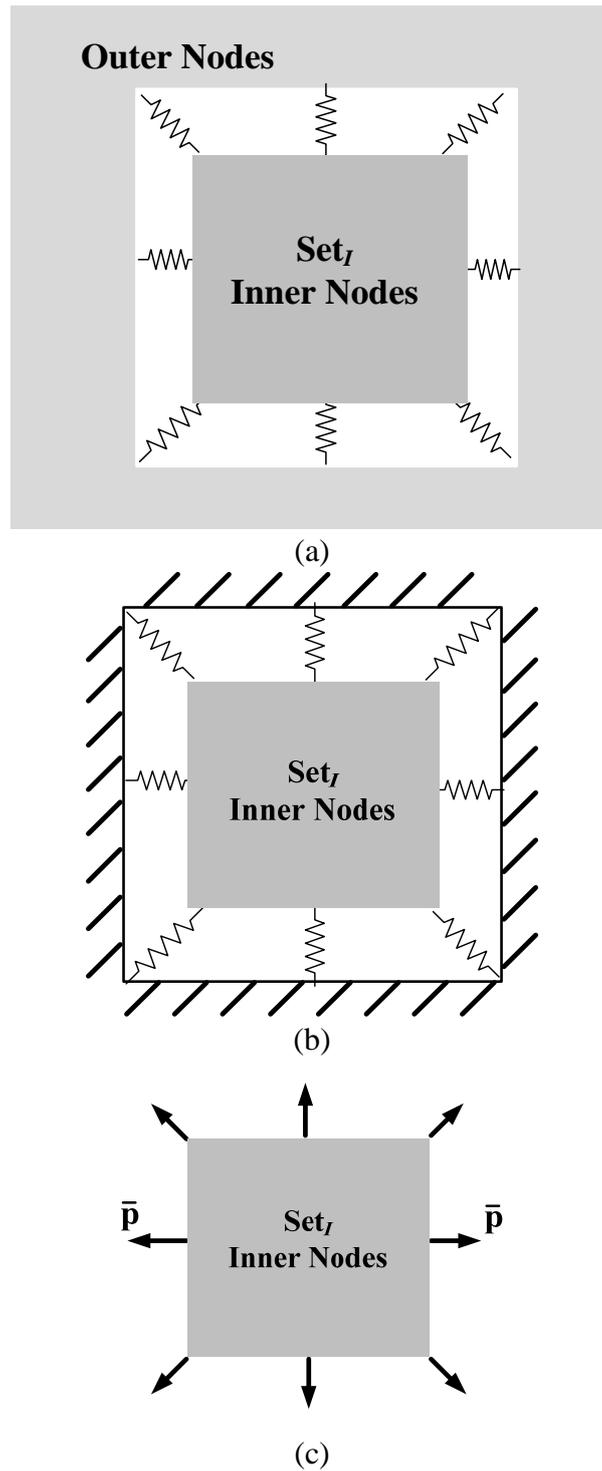

Fig. 2 (a) Schematic diagram of inner nodes and the outer nodes of $Set_I$, (b) local displacement-controlled relaxation and (c) local force-controlled relaxation.

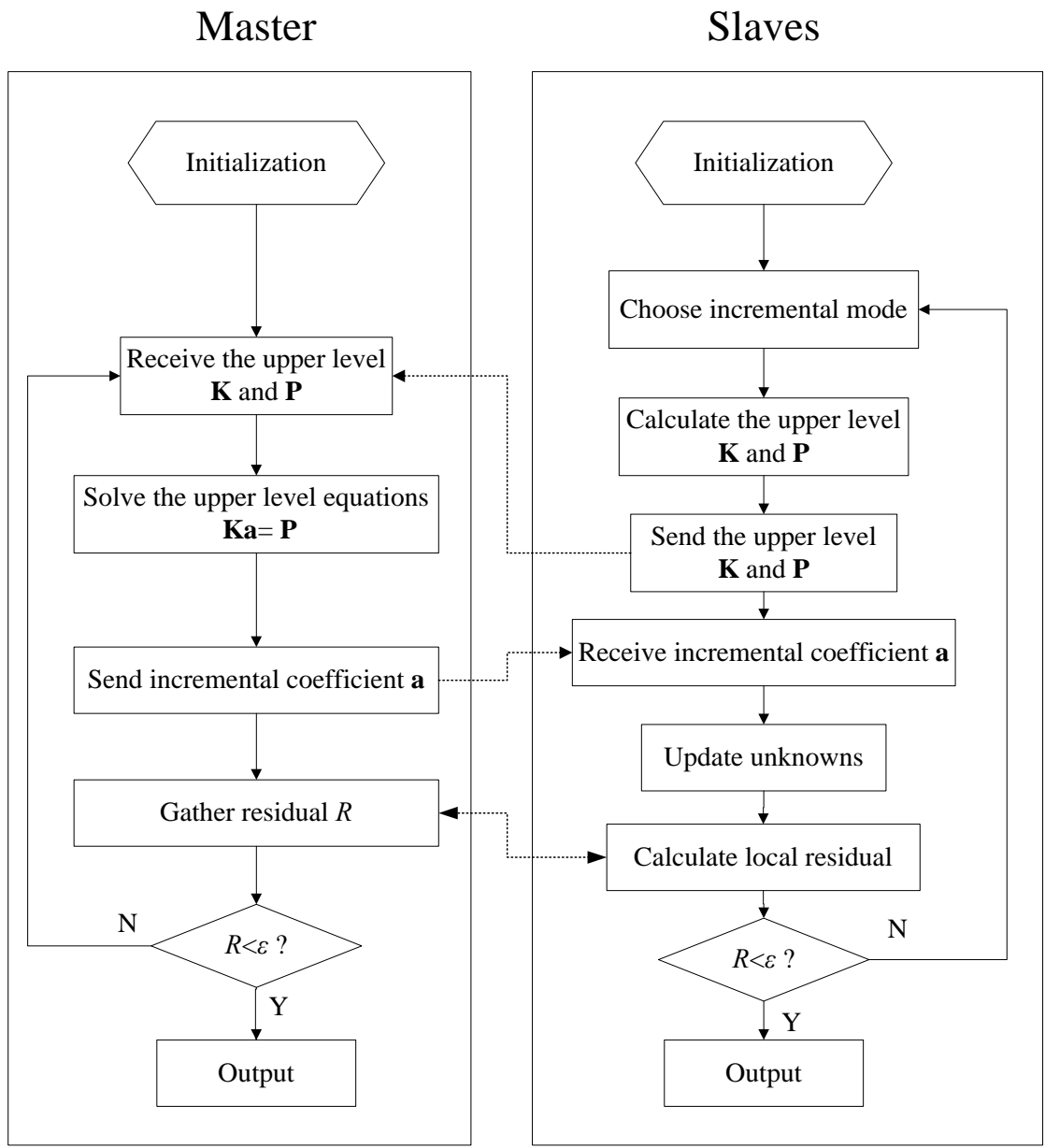

Fig. 3　Flowchart showing the hierarchical parallel algorithm.

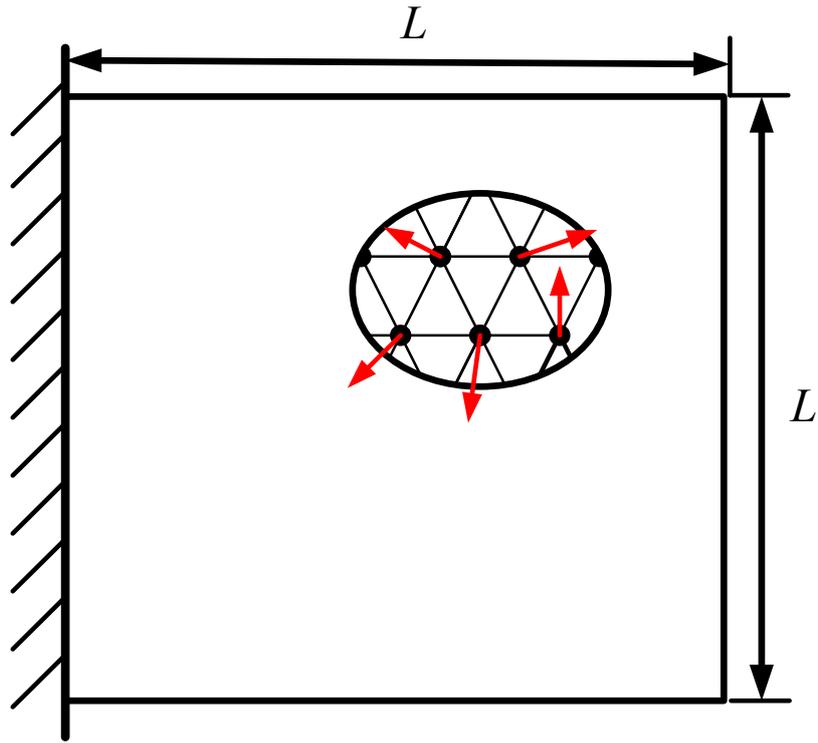

Fig. 4 Two-dimensional square truss system with random force on each node.

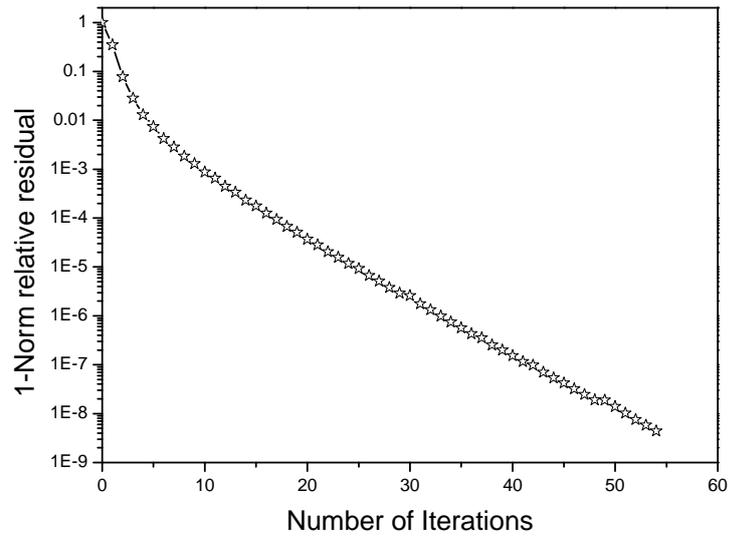

Fig. 5  Relative residual of super large one billion DOFs problem decreases as a quasi- exponentially function of iterations number. The problem is divided into 2000 sets.

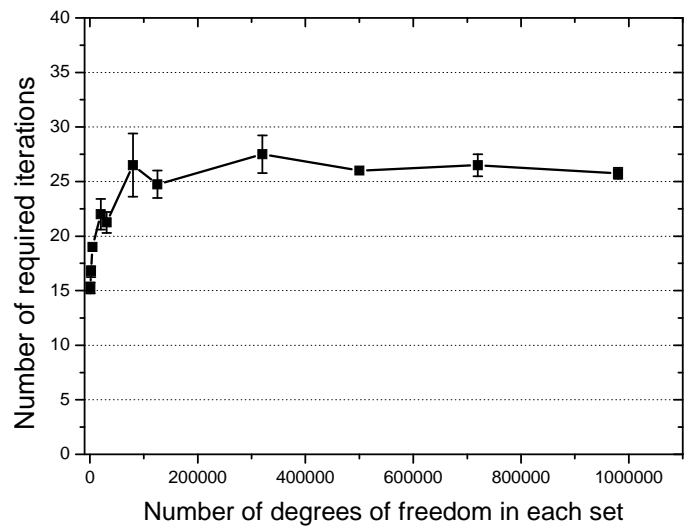

Fig. 6 The number of iteration required (NIR) to meet specific accuracy requirement ($5\times10^{-6}$) versus the number of DOFs in each set: a 2-dimension square system with 64 sets are tested with four different random loads.

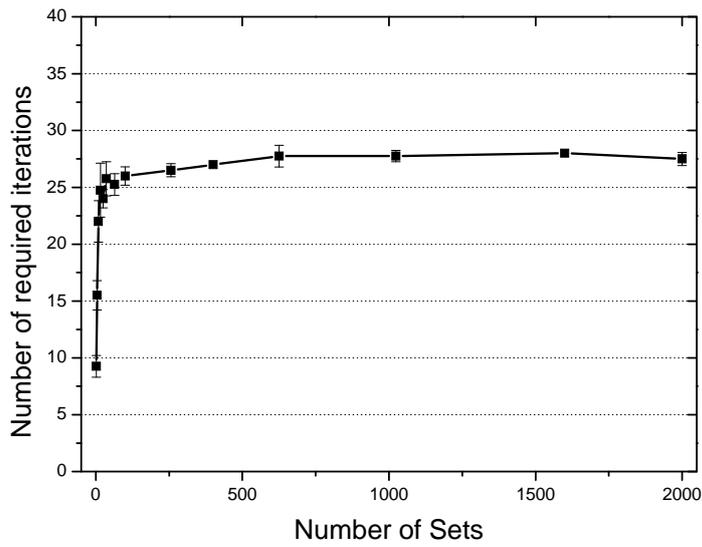

(a)

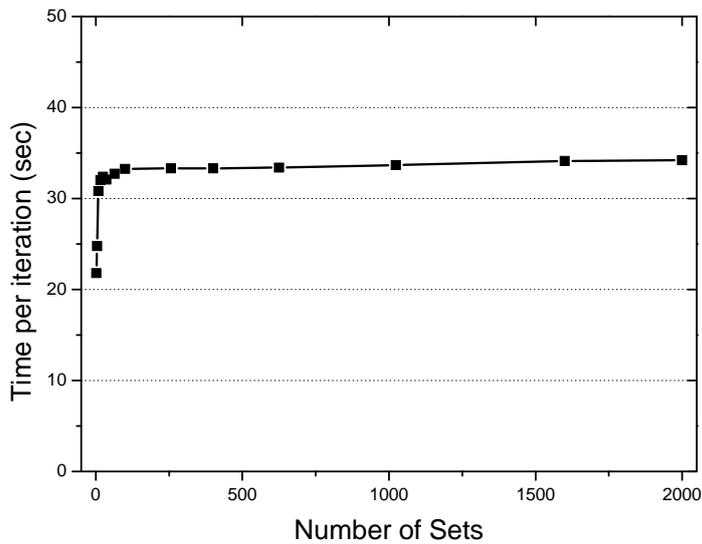

(b)

Fig. 7 The 2-dimension square system is tested with 2-2000 sets, and each set has half a million DOFs. The number of iteration required (NIR) to meet specific accuracy requirement ($5\times10^{-6}$) and elapsed time per iteration is plotted as a function of set number in (a) and (b), respectively. NIR shows a rapid convergence and the elapsed time per iteration presents a very slow growth rate with increasing sets.

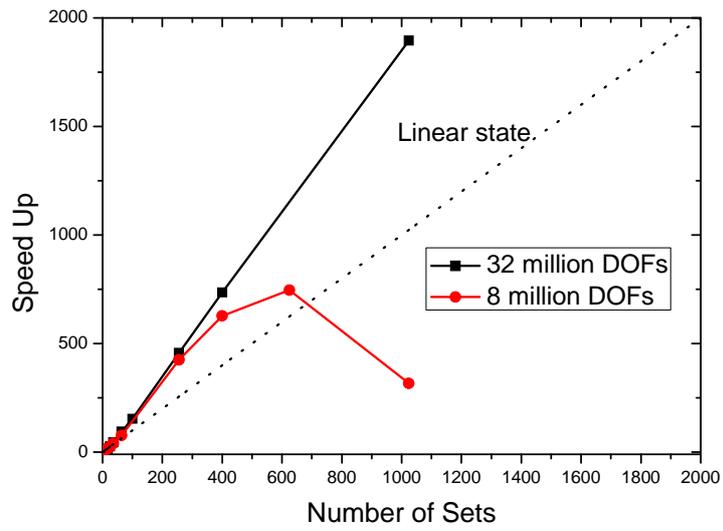

Fig. 8 The parallel improvement of our algorithm is tested with 8 million DOFs and 32 millions DOFs examples.

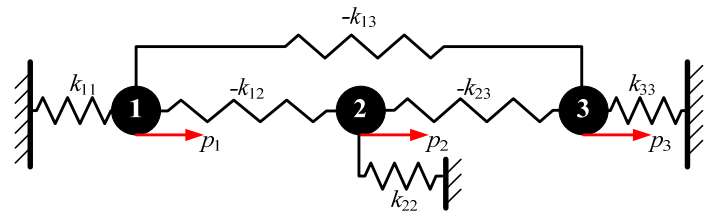

Fig. 9  A one-dimension spring system is used to represent the generally symmetric sparse linear system.

Table 1 Specific details on "Exploration 100" cluster

| Device Parameters | | | |
|---|---|---|---|
| Computing nodes | 740 | Storage nodes | 22 |
| O.S. | Red Hat® 4.1.2-48 | CPU | Intel® E5670 |
| Ram | 32G DDR-1600 | Cores/CPU | 12/2 |
| Complier | Intel® Fortran 11.1069 | MPI | Intel® MPI |